% MNSAMPLE.TEX
%
% A sample plain TeX single/two column Monthly Notices article.
%
% v1.5  --- released 25th August 1994 (M. Reed)
% v1.4  --- released 22nd February 1994
% v1.3  --- released  8th December 1992
%
% Copyright Cambridge University Press

% The following line automatically loads the mn macros if you are not
% using a format file.
\ifx\mnmacrosloaded\undefined \input mn\fi

% If your system has the AMS fonts version 2.0 installed, MN.tex can be
% made to use them by uncommenting the line: %\AMStwofontstrue
%
% By doing this, you will be able to obtain upright Greek characters.
% e.g. \umu, \upi etc.  See the section on "Upright Greek characters" in
% this guide for further information.

\newif\ifAMStwofonts
%\AMStwofontstrue

\ifCUPmtplainloaded \else
  \NewTextAlphabet{textbfit} {cmbxti10} {}
  \NewTextAlphabet{textbfss} {cmssbx10} {}
  \NewMathAlphabet{mathbfit} {cmbxti10} {} % for math mode
  \NewMathAlphabet{mathbfss} {cmssbx10} {} %  "   "    "
  \ifAMStwofonts
    \NewSymbolFont{upmath} {eurm10}
    \NewSymbolFont{AMSa} {msam10}
    \NewMathSymbol{\upi}     {0}{upmath}{19}
    \NewMathSymbol{\umu}     {0}{upmath}{16}
    \NewMathSymbol{\upartial}{0}{upmath}{40}
    \NewMathSymbol{\leqslant}{3}{AMSa}{36}
    \NewMathSymbol{\geqslant}{3}{AMSa}{3E}

     \let\ge=\geqslant
  \else
    \def\umu{\mu}
    \def\upi{\pi}
    \def\upartial{\partial}
  \fi
\fi

% Marginal adjustments using \pageoffset maybe required when printing
% proofs on a Laserprinter (this is usually not needed).
% Syntax: \pageoffset{ +/- hor. offset}{ +/- vert. offset}
% e.g.    \pageoffset{-3pc}{-4pc}

\pageoffset{-2.5pc}{0pc}

\loadboldmathnames

%\Referee   %  uncomment this for referee mode (double spaced)

% \pagerange, \pubyear and \volume are defined at the Journals office and
% not by an author.

% \onecolumn        % enable one column mode
% \letters          % for `letters' articles
\pagerange{1--4}    % `letters' articles should use \pagerange{Ln--Ln}
%\pubyear{1989}
%\volume{226}
% \microfiche{}     % for articles with microfiche
\authorcomment{}  % author comment for footline

\begintopmatter  %  start the two spanning material

\title{Emergence of magnetic field due to spin-polarized baryon matter in neutron stars}
\author{M. Kutschera}
\affiliation{Astrophysics Division, H. Niewodnicza\'nski
Institute of Nuclear Physics, ul. Radzikowskiego 152, 31-342 Krak\'ow, Poland}

\affiliation{and}

\affiliation{Institute of Physics, Jagiellonian University, ul.
Reymonta 4, 30-059 Krak\'ow, Poland}

\shortauthor{M. Kutschera}
\shorttitle{Emergence of magnetic field due to spin-polarized baryon matter in neutron stars}

% \acceptedline is to be defined at the Journals office and not
% by an author.

\acceptedline{Accepted 1999 March 9. Received 1999 March 5;
  in original form 1998 August 14}

\abstract {A model of the ferromagnetic origin of magnetic fields of
neutron stars is  considered. In this model, the
magnetic phase transition occurs inside the core of neutron stars
soon after formation. However, owing to the high electrical
conductivity  the core magnetic field is
initially fully screened. We study how this magnetic field
emerges for an outside observer.
After some time, the induced field that  
screens the ferromagnetic field decays enough to uncover a detectable fraction 
of the ferromagnetic field. 
We calculate the time-scale of decay of the screening field and study
how it depends on the size of the ferromagnetic core. We find that the same
fractional decay of the screening field occurs earlier for larger cores.
We conjecture that weak fields of millisecond 
pulsars, $B \sim 10^8 - 10 ^9 G$, could be identified with ferromagnetic fields of
unshielded fraction $\epsilon \sim 10^{-4}- 10^{-3}$ resulting
from the decay of screening fields
by a factor ${\it 1-\epsilon}$ in $\sim 10^8$ yr since their birth.}


\maketitle  %  finish the two spanning material

\section{Introduction}

The physical origin of the magnetic field of neutron stars remains 
an open problem. Some researches adopt a working hypothesis that 
the field is inherited from the progenitor star. There also exist
more specific models of generation of the magnetic field during
the early cooling
of a young neutron star. Generally, this kind of approach
links the magnetic field of neutron stars to macroscopic
currents generated either by the motion of the ionized material or by heat
transport (Chanmugam 1992).

For neutron stars, unlike for any other
astrophysical object, there exists a different physical 
possibility, namely that the magnetic field is of microscopic origin and
reflects spin-polarization properties of the neutron star matter itself. 
The ground state of dense baryonic matter is determined by nuclear
interactions that possess very strong spin components. It seems likely that these
interactions lead to spin ordering in the neutron star matter
making the ground state of the system
a permanently magnetized phase, at least in some
range of densities. 

Mechanisms of spin ordering in dense matter have been studied by a number of authors 
(Kutschera \& W\'ojcik 1989, 1990, 1993,  1996, 1997; 
Kutschera, Broniowski \& Kotlorz 1990a,b; Niembro et al. 1990;
Marcos et al. 1991; Kutschera 1994a,b).
Kutschera \& W\'ojcik (1989) have shown that neutron star matter
with a low
proton fraction of a few per cent is particularly susceptible to 
ferromagnetic spin ordering.
We have shown (Kutschera \& W\'ojcik 1990, 1993, 1995; Kutschera
1994a,b) that protons in neutron star matter with such a
small  
proton admixture behave as polarons that can localize at higher 
densities. Neutron star matter with localized protons displays a 
ferromagnetic instability (Kutschera \& W\'ojcik 1989,
1996,1997) and the spin-ordered system develops a permanent 
magnetization,
$M \sim n_P \mu_{eff}$, where $n_P$ is the proton number
density
and $\mu_{eff}$ is the effective magnetic moment of the localized proton. 
Typical values of magnetization are in the range $10^{13}-10^{14} G$ (Kutschera \& W\'ojcik 1996). 

There exist also other model mechanisms
that result in a spontaneous magnetization of hadronic matter
inside neutron stars. The presence of the neutral pion
condensate in the nucleon matter can lead to  ordering of nucleon spins (Dautry \& Nyman 1979). 
The same mechanism can operate in  quark matter with chiral
condensate (Kutschera et al. 1990a,b). The magnetization can be higher in this case, up to $\sim 10^{15} G $.

Uncertainties in the physics of dense hadronic matter prevent a firm conclusion concerning
the nature of the ground state of baryon matter in neutron stars. However,
ubiquity of spin interactions between hadrons with strong couplings makes 
the spin-polarized ground state of dense baryon matter a
possibility that deserves serious consideration. 

Independently of
the mechanism of polarization, one can study the phenomenological consequences of
the existence of spin-polarized baryon matter in neutron stars.
If the ground state of dense matter is permanently magnetized at
baryon number densities 
exceeding some critical density $n_f$, then there could exist a class of 
neutron stars  possessing magnetic
cores provided $n_f<n_c^{max}$, where $n_c^{max}$ is the central baryon density of
the maximum mass neutron star. Only sufficiently massive stars, with central 
densities $n_c>n_f$ would belong to this class. The critical density, $n_f$, is estimated
to exceed twice the nuclear saturation density. The magnetized core is a source of the 
magnetic field. However, the phase transition producing the magnetized core
occurs in the inner part of the star, which has very high electrical
conductivity. Thus, any magnetic 
field due to the sudden appearance of the spontaneous polarization of baryonic matter becomes fully 
screened in such a medium. It takes some time for the magnetized core to become
detectable by an outside observer. The aim of 
this paper is to investigate this problem. 

The fact that any magnetic moment spontaneously created in the
neutron star core is initially fully screened has important
consequences for the spin structure of the magnetic core, which
can form a single magnetic domain. Such a domain is the most
favourable configuration as far as the hadronic energy gain
resulting from spin
ordering is concerned. Because of screening, electromagnetic
interactions do not force the single domain 
core to fragment into smaller magnetic domains.

In the next section  we describe the model of the magnetized core. In Section 3 we study the
screening of the ferromagnetic field. Then in Section 4 the decay of the screening currents
is considered. Section 5 contains astrophysical implications of the model. Some useful formulae
are given in Appendix A.

\section{The magnetized core inside a neutron star}

A single domain structure of the magnetic core is favoured by
the dynamics of the magnetic phase transition.
As the neutron star matter cools,
the conditions for the spontaneous spin ordering to occur are met for the first
time at the centre of the star. This is because the
energy gain resulting from spin ordering in the ferromagnetic
phase increases with density (Kutschera \& W\'ojcik 1990, 1993,
1995, 1996). Thus, additionally, the critical temperature
increases with density, and, for a roughly isothermal core, the
first small bubble of the spin-ordered phase is nucleated at the
centre of the
neutron star. This first small domain 
grows as the temperature drops with nucleon spins in the next
layers of matter being polarized 
in the same direction. A single domain is the minimum hadronic
energy configuration of the spin-polarized phase of neutron star matter.
So far we have neglected electromagnetic interactions in the
polarized phase. One can worry that, when
magnetostatic energy is included, such a
single domain may fragment into smaller randomly oriented
domains. We show at the end of Section 3 that this is not the case.

To model the magnetized core we assume that the spontaneously polarized matter
forms a single domain with magnetization in the z-direction:
${\bf M}=M(r) {\hat {\bf z}} $. For this study the only relevant quantity is 
the magnetization as a function of baryon number density, $M\equiv M(n_B)$.
Calculations reported by Kutschera \& W\'ojcik (1996) show that the 
magnetization depends only weakly on density, $M\approx M_0 \sim 10^{13}-10^{14} G$
in the whole range of relevant densities.
In the case of polarized quark matter with  chiral condensate, the
magnetization is  essentially density independent (Kutschera et al. 1990a,b).
We can thus parametrize the magnetization distribution inside neutron stars
as
$$
M(r)=M_0 f(r), \eqno\stepeq
$$
where the function $f(r)$ accounts for the radial variation of magnetization. 
In the following we adopt a simple form: 
$ f(r)=1$ for $ r<r_a$ and $f(r)=0$ for $r>r_b$. This 
function corresponds to a uniformly magnetized sphere of radius $r_a$  with a
surface layer of thickness $h=r_b-r_a$ where the magnetization
drops linearly
to zero at $r=r_b$ with $f(r)=(r-r_b)/(r_a-r_b)$ (Fig.1). This simple function 
is flexible enough to account for the main features of the
magnetization distribution 
in neutron stars.

\beginfigure{1}
%\vskip 91mm
\vskip 21mm
\caption{{\bf Figure 1.} The magnetization profile function f(x)
and the function g(x) for $z_b=0.9$. Solid and dashed curves correspond, respectively,
to $z_a=0.001$ and $z_a=0.8$.}

\endfigure

The magnetic moment of the core is
$$
 d_{core}= \int_0^{r_b} M(r) d^3r, \eqno\stepeq
$$
with $M(r)$ given by equation (1).
The size of the core is
determined by the critical density for spontaneous polarization,
$n_f$. 
Thus the nonzero magnetization, $M(r) \ne 0$, exists only in the
inner core of stars of density $n_B(r)>n_f$. The core radius $r_b$ corresponds
to the critical density $n_f$: $n_B(r_b)=n_f$.

The contribution of the magnetic moment (2) to the magnetic field at the 
pole is  
$$
 B_P^{fer} = {2d_{core} \over R_{NS}^3}, \eqno\stepeq
$$
where $R_{NS}$ is the radius of the neutron star. This field 
is $B_P \sim 10^{12} - 10^{13}$ G for the above typical values of
the magnetization.

\section{Screening of the ferromagnetic field}

Calculations in various models suggest the energy per baryon in
the polarized phase could be below that for the normal phase by at least $\sim 1MeV$.
The phase transition from normal matter, with no spin ordering, to
magnetized matter is thus expected to occur very soon after the
formation of the neutron star. The phase transition is 
completed quickly. For any practical purposes one can assume that 
the ferromagnetic field is switched on instantaneously.
This is because the time-scale for magnetic field diffusion, determined 
by the electrical conductivity, is much longer than the duration of the phase transition.

The neutron star matter outside the magnetic core is a medium of very
high electrical conductivity $\sigma$. Realistic calculations
show that conductivities
corresponding to the neutron star crust are, typically,
$\sigma_{crust} \sim 10^{23} s^{-1}$ (Chanmugam 1992). These values are
lower than those corresponding to the core, 
$\sigma_{core} \sim 10^{29} s^{-1}$ (Chanmugam 1992). It is rather  obvious that sudden
switching-on of the magnetic field of the magnetized core will
result in the induction of the screening field, which will
fully shield the ferromagnetic field. This happens because for such a high 
conductivity the magnetic flux through any loop is conserved on
the time-scale of the phase transition. Flux
conservation then requires that ${\partial {\bf B}}/{\partial t}=0$. We
assume that there is no magnetic field  before the phase
transition occurs, ${\bf B}=0$. Thus, at the instant $t=0$ the switched-on
ferromagnetic field, ${\bf B}_{fer}$, and the induced field,
${\bf B}_{ind}(t)$, cancel one another exactly,
$$
{\bf B}_{fer} + {\bf B}_{ind}(0) =0. \eqno\stepeq
$$

The ferromagnetic field of the core is easily calculated. The
vector potential has only the $\phi$-component 
$$
A_{\phi}={R(r) \over r} sin\theta, \eqno\stepeq 
$$
where the function $R(r)$ reads
$$ 
R(r)=d(r)/r. \eqno\stepeq 
$$
 Here $d(r)$ is the magnetic moment inside the sphere of
radius $r$,
$$ 
d(r)= \int_0^{r} M(r') d^3r'. \eqno\stepeq
$$

Components of the magnetic field are
$$ 
B^{fer}_{r}={2 R(r) \over  r^2} cos \theta, \eqno\stepeq 
$$

$$ 
B^{fer}_{\theta}=-{1 \over r}{{\partial R(r)} \over {\partial r}}
sin \theta. \eqno\stepeq 
$$

The flux conservation condition, equation (4), gives the components of
the induced field that screens the ferromagnetic field,
$$
B^{ind}_{r}(0)=-B^{fer}_{r}, \eqno\stepeq
$$

$$ 
B^{ind}_{\theta}(0)=-B^{fer}_{\theta}. \eqno\stepeq
$$
These formulae show that the induced field components,
$B^{ind}_{r}$ and $B^{ind}_{\theta}$, satisfy equations (8) and (9) with
the function $R_{ind}(r,t)$, which, however, depends also on time. At $t=0$ 
the condition (4) implies that
$$
R_{ind}(r,0)=-R(r). \eqno\stepeq
$$

One can also obtain the space structure of the current
sustaining the induced field. Generally, this current changes in time, 
${\bf J}_{ind} \equiv {\bf J}_{ind}(r,\theta,t)$. 
From the formula $\nabla \times {\bf B}_{ind}=4\pi /c {\bf J}_{ind}$ it follows that
$$ 
{4\pi \over c}J^{ind}_{\phi}={sin\theta \over
r}({\partial^2 R_{ind} \over {\partial r^2}}-{2R_{ind} \over r^2}). \eqno\stepeq
$$
At $t=0$ using equation (4), we find
$$ 
{\bf J}_{ind}(r,\theta,0)=-c \nabla \times {\bf M}. \eqno\stepeq
$$

The above discussion allows us now to address the question of
the stability of a large single domain against fragmentation into
smaller randomly oriented domains. In the case of terrestrial
ferromagnets, large domains are not stable. Their magnetostatic
energy can be lowered by producing a number of smaller domains
with apparently randomly oriented magnetic moments. Let us
stress that
reduction of the energy is due to dipole magnetic
interactions between these magnetic moments.
The magnetic core in neutron stars forming  a single domain
with screened magnetic field cannot lower its magnetostatic
energy by fragmenting into 
smaller domains. This is because the magnetic moment of any
new domain would also be fully screened and thus, unlike in
terrestrial ferromagnets, there would be no magnetic dipole
interactions between neighbouring domains. The core is thus
stable against fragmentation into small domains.

\section{Decay of screening fields}

The induced current, ${\bf J}_{ind}$, will suffer  ohmic decay
as the electrical  
conductivity, $\sigma$, though very high, is finite. 
The nonzero net field, ${\bf B}={\bf B}_{fer}+{\bf B}_{ind}(t)\ne 0$, will eventually
emerge. Let us note that ohmic decay is the only relevant mechanism of magnetic
field decay as long as the net magnetic field is low. Ambipolar diffusion and
Hall drift, which play a crucial role in the decay of strong
fields (Goldreich \& Reisenegger 1992),  can be safely neglected.

To calculate the time behaviour of the screening field we apply a standard
analysis of decay modes (Wendell, Van Horn \& Sargent 1987). To
avoid unnecessary complications in our exploratory 
study we neglect 
spatial and temporal dependence of the conductivity and assume
that $\sigma=const$.  
The time dependence of the induced field can
be found from the expansion of the function $R_{ind}(r,t)$ into eigenfunctions 
$$
X_n(x)={\sqrt 2} n\pi x j_1(n\pi x), \eqno\stepeq
$$
where $j_1$ is the Bessel function and $x=r/R_*$ is the normalized radial 
variable with a suitably chosen radius $R_*$. The expansion reads
$$
R_{ind}(x,t)=\sum_n C_n X_n(x) \exp{(-t/\tau_n)}, \eqno\stepeq
$$
where $\tau_n=4\pi R_*^2\sigma/(c\pi n)^2$ is the decay time of the $n$th mode.

The expansion coefficients, $C_n$, are obtained using the function
$R_{ind}(x,0)=-R(x)$ as the initial condition:
$$
C_n= \int_0^1 R_{ind}(x,0) X_n(x) dx. \eqno\stepeq
$$

With our choice of the magnetization profile, $f(r)$, the function $R(x)$
reads
$$
R(x)={4 \over 3} \pi M_0 R_*^2 g(x), \eqno\stepeq
$$
where the function $g(x)$ is determined entirely by the magnetization profile
function $f(r)$. The function $g(x)$,  calculated by integrating the 
magnetization in equation (7), is given in Appendix A.

The form (18) of the function $R(x)$ indicates that the time behaviour of the
screening field is sensitive to the magnetization profile. To 
study this behaviour we have calculated the coefficients $C_n$ of the 
expansion, equation (17), which are also given in Appendix A.
The unshielded fraction, $\epsilon(x,t)$,   
of the ferromagnetic field emerging after time $t$ is
$$
\epsilon(x,t)={{\vert {\bf B}_{fer}+{\bf B}_{ind}(t) \vert} \over {\vert {\bf B}_{fer} \vert}}
=1-{R_{ind}(x,t) \over R_{ind}(x,0)}. \eqno\stepeq
$$
We are interested mostly in the value of $\epsilon(x,t)$ at the
surface of the star, 
$x=1$, which we denote as $\epsilon(t) \equiv \epsilon(1,t)$. 

The formula (19) shows that $\epsilon(t)$ does not depend on the 
magnetization $M_0$. This means that the relative rate of emergence of the 
ferromagnetic field depends only on the geometry of the magnetized core.
For the case of uniform conductivity we consider here this behaviour is even
more universal. We can express the time behaviour of the emerging magnetic 
field in terms of the dimensionless variable $y=t/\tau_1$, where
$$
\tau_1={{4 R_*^2 \sigma} \over c^2 \pi} \eqno\stepeq
$$
is the decay time of the longest-living (fundamental) decay mode. The formula 
(19) becomes
$$
\epsilon(y)=1-{\sum_n C_n X_n(1)\exp{(-yn^2)} \over \sum_n C_n X_n(1)}. \eqno\stepeq
$$

The main results of our analysis are presented in Fig.2 where we show $\epsilon(y)$
for indicated values of the parameters $z_a\equiv r_a/R_*$ and
$z_b \equiv r_b/R_*$ governing the spatial distribution of
magnetization  in our simple model. As one can notice, the shape of 
the curves is determined essentially by the value of $z_b$ with much smaller 
influence of  $z_a$.

\beginfigure{2}
%\vskip 91mm
\vskip 21mm
\caption{{\bf Figure 2.} The unshielded fraction $\epsilon$ as a function
of the variable $y=t/\tau_1$ for indicated values of the magnetized core
parameters $z_a$ and $z_b$. Curves labelled 1,2 and 3 correspond to $z_b \approx 0.99$,
$z_b \approx 0.9$ and $z_b \approx 0.4$, respectively.}

\endfigure
The general tendency is that the higher is $z_b$ the higher is $\epsilon$. The 
curves for the same $z_b$ and various $z_a$ are very close to each other.
We see that the time dependence of the unshielded fraction $\epsilon$ is most
sensitive to the total radial extension of the ferromagnetic core. The radial
distribution of magnetization, which is our model is controlled by $z_a$, has
a much smaller influence on $\epsilon$.
This means that the ferromagnetic field for cores of larger extension becomes 
uncovered earlier than for smaller ones. The curves shown in Fig.2 prove that
the unshielded fraction $\epsilon$ is very sensitive to $z_b$ and is much
less sensitive to $z_a$. For each value of $z_b$ we show curves corresponding 
to two extreme values of $z_a$: $z_a=0.001$ (dashed curves) and
$z_a=0.9z_b$ (solid curves).

In our discussion above we have made a number of simplifying assumptions regarding
the neutron star structure. The conductivity is assumed to be
uniform and  constant in time, 
which is equivalent to constant temperature. Also, Euclidean geometry is used.
These simplifications are not
expected to affect our qualitative conclusions as far as the decay time-scale of 
the screening field is concerned.

\section{Conclusions and implications}

The main conclusion from the above analysis is that the emergence time of the
ferromagnetic field, at low values of the unshielded fraction $\epsilon$, is 
very sensitive to the size of the polarized core. For large cores, comprising
$\sim 90$ per cent of the radius, a fraction $\epsilon \sim
10^{-4}$ of the core field is 
visible after $t \sim 10^{-4} \tau_1$. For small cores, $z_b \sim 0.4$, the same fractional
field is visible after $t \sim 10^{-1} \tau_1$, i.e. a factor of $\sim 10^3$ later.

To explore the astrophysical consequences of this model we must
specify the relevant values of the decay time $\tau_1$, or equivalently the
electrical conductivity $\sigma$. For magnetized cores in neutron stars, the
relevant conductivity is that of the liquid core matter, $\sigma_{core} \sim 10^{29} s^{-1}$.
The corresponding decay time is $\tau_1 \sim 10^{12}$ yr. One should notice
that the choice of the $\sigma_{core}$ implies that the presence of the neutron
star crust becomes somewhat irrelevant for our considerations here. This is because
the crust conductivity is lower by a factor $\sim 10^{-6}$. An important consequence
is that the value of the radius $R_*$ we use in the definition
of the variable $x$ 
should be identified with the radius of the liquid core,
$R_*=R_{lc}$, rather than the total neutron 
star radius (which includes the crust). Correspondingly, $z_a$
and $z_b$ are the 
fractions of the liquid core radius, $z_a=r_a/R_{lc}$ and $z_b=r_b/R_{lc}$.
With this observation the values $z_b \ge 0.9$ do not seem unrealistic, especially
for soft equations of state.

Applying our model to neutron stars we find that the magnetic
field of the magnetized  
core formed soon after the birth of the star emerges in $\sim 10^{8}$
yr at a level 
of $\sim 10^{-4} B^{fer}_P \sim 10^{8} G$ provided the
ferromagnetic core is large enough. 
Both the time-scale of  $10^8$ yr and the magnetic field  of $10^8 G$ are 
typical for millisecond pulsars. One can thus conjecture that the magnetic fields
of millisecond pulsars are due to spin-polarized matter inside neutron stars.

The ferromagnetic origin of the magnetic fields of millisecond pulsars could explain
the discrepancy between the birth rate of low mass X-ray
binaries (LMXB), which are 
supposed to be the progenitors of millisecond pulsars, and the number of 
millisecond pulsars in the Galaxy (Bhattacharya 1995). In this
case, single neutron stars born with  
short rotation periods and low magnetic fields (much less than $10^7 G$) could
become millisecond pulsars in $\sim 10^8$ yr. This would also help to
understand the existence of single millisecond pulsars without invoking the
companion evaporation scenario. 
Also, this model would shed some new light on the old question
of the decay time of  
the inherited magnetic field of normal radio pulsars. Currently,
a popular view is 
that only magnetic fields of neutron stars that accreted matter
in binary evolution decay 
significantly. In the ferromagnetic model, the inherited
magnetic field of an isolated neutron star could decay in 
$\sim 10^7$ yr, as was found in early studies of the subject
(Ostriker \& Gunn 1969). There would 
be no need for direct connection between the amount of matter accreted by the 
neutron star in LMXBs and the amount of decay of its magnetic
field. Recent analysis by Wijers (1997)  
provides some evidence against field decay being proportional to the
mass accreted by 
a neutron  star during the LMXB evolution, supporting our conjecture.

Let us stress finally that looking for the presence (or absence)
of a ferromagnetic component of 
the magnetic field of neutron stars is of great importance for
the physics of
hadronic matter. Physical conditions prevailing in neutron stars (high density, low temperature and $\beta-equilibrium$)
are not accessible to 
any laboratory experiment. Magnetic field could serve as a direct probe of dense
hadronic matter in the neutron star interior. Evidence in favour
of the presence  of spin-polarized matter inside neutron stars
would introduce new qualitative features of the ground state 
of dense hadronic matter. An opposite conclusion would mainly constrain the critical
density $n_f$. The fact that a magnetic field resulting from the
ferromagnetic core 
grows with time, while the inherited field decays in time, could
potentially allow one  
to distinguish the two components.

\section*{Acknowledgements}
This research was partially supported by the Polish State Committee for Scientific
Research (KBN), under Grant 2 P03D 001 09.

\appendix

\section{~}

For our choice of the magnetization profile, $f(r)$, the function $g(x)$ is a
smooth function, which is given analytically below:

$$
g(x)= x^2,~~~~~ x<z_a, \eqno\stepeq
$$

$$
g(x)={A \over x}+Bx^2+Cx^3, ~~~~~ z_a<x<z_b, \eqno\stepeq
$$
where
$$
A=z_a^3+{{0.75z_a^4-z_bz_a^3} \over {z_b-z_a}}, \eqno\stepeq
$$

$$
B={z_b \over {z_b-z_a}}, \eqno\stepeq
$$
and, 
$$
C={-0.75 \over {z_b-z_a}}, \eqno\stepeq
$$
and, finally,
$$
g(x)={D \over x},~~~~~~z_b<x<1, \eqno\stepeq
$$
where
$$
D=z_a^3+{1 \over {z_b-z_a}}({3 \over 4}(z_a^4-z_b^4)+z_b(z_b^3-z_a^3)).\eqno\stepeq
$$

The expansion coefficients $C_n$ in equation (19) read:
$$
C_n=-{4 \over 3} \pi M_0 R_*^2 c_n, \eqno\stepeq
$$
where
$$
c_n=2[a_n(z_a)+A(b_n(z_a)-b_n(z_b))+B(a_n(z_b)-a_n(z_a))
$$

$$
+ {C \over (n \pi)^4}
(d_n(z_b)-d_n(z_a))+Db_n(z_b)]. \eqno\stepeq
$$
Here
$$
a_n(z)={-3n \pi z cos(n \pi z)+3sin(n \pi z) \over (n \pi)^3}
$$

$$
-{(n \pi z)^2sin
(n \pi z) \over (n \pi)^3}, \eqno\stepeq
$$

$$
b_n(z)={sin(n \pi z) \over n \pi z}, \eqno\stepeq
$$
and
$$
d_n(z)=8cos(n \pi z)-4(n \pi z)^2cos(n \pi z)+8n \pi z sin(n \pi z)
$$

$$
-(n \pi z)^3
sin(n \pi z). \eqno\stepeq
$$

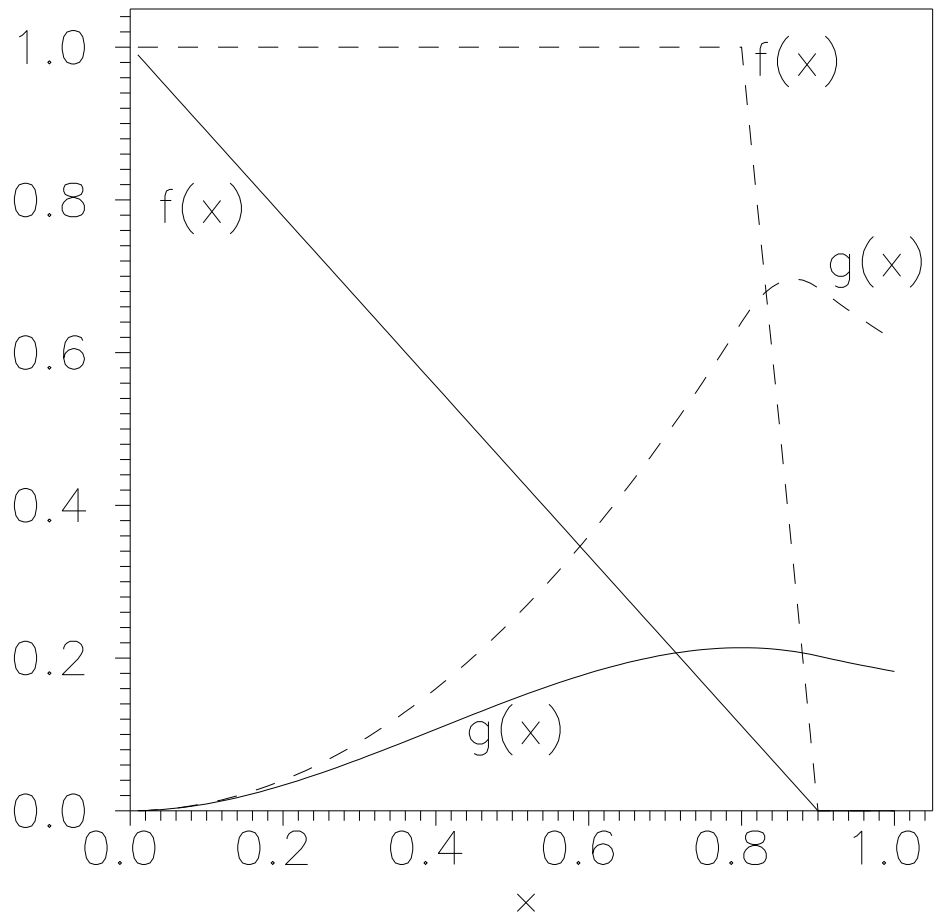
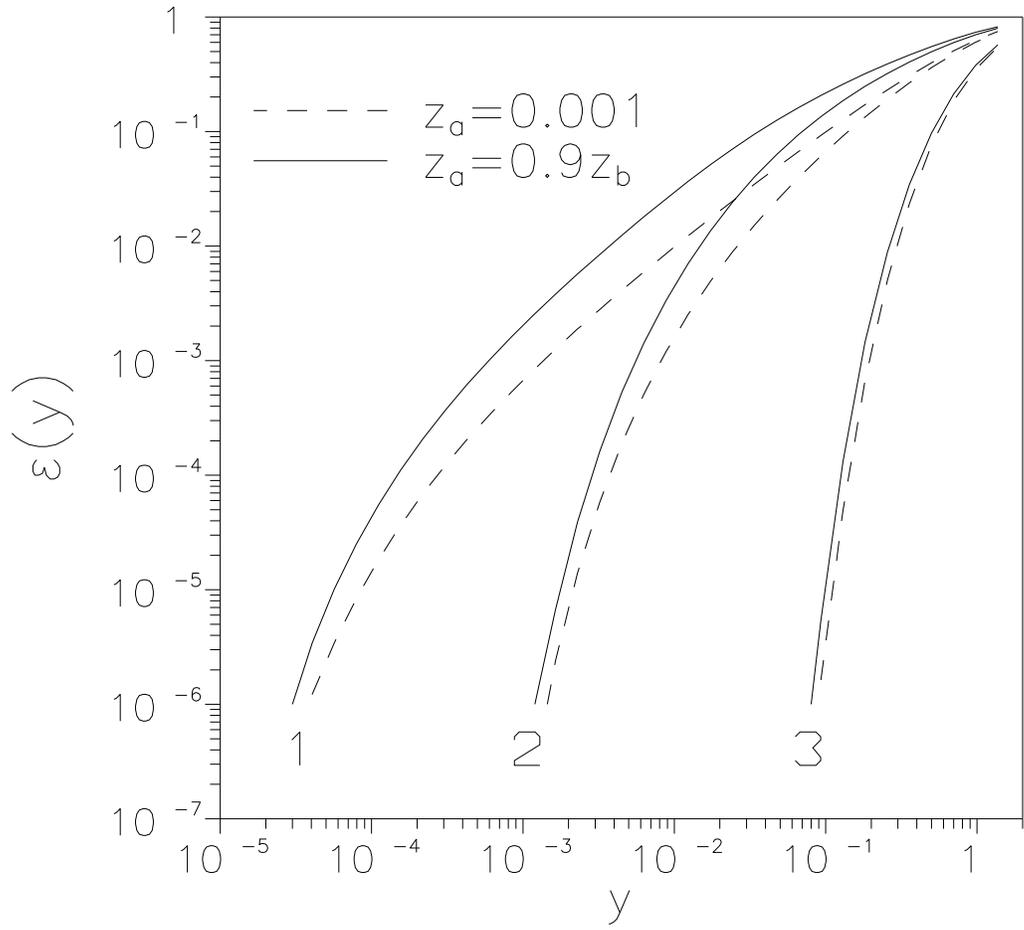
\bye